\documentclass{article}
\usepackage{slashed}
\begin{document}
\title{On binary pulsars and the force of gravity}
\author{Davor Palle \\
ul. Ljudevita Gaja 35, 10000 Zagreb, Croatia \\
email: davor.palle@gmail.com}
\maketitle

\begin{abstract}
The energy-momentum budget of the astrophysical systems can be 
studied by the exact local conservation equation derived
by Landau and Lifshitz. We show that a similar equation
is valid for the Einstein-Cartan gravity.
We reanalyze a binary pulsar system using the Landau-Lifshitz conservation
equation and show that the orbital
period change rate can be completely understood
as a curvature backreaction process.
Taking into account the detailed theoretical and observational research
of relativistic binary pulsar systems, especially the system
of Hulse and Taylor, we conclude that general
relativity and astrophysical observations rule out the existence 
of gravitational radiation. 
We comment upon the LIGO GW events and their alternative explanation, 
as well as the recent pulsar timing arrays data.
\end{abstract}
\vspace{6mm}

\section{Introduction and motivation}

The force of gravity represents for physicists the most intriguing and challenging 
fundamental interaction to be studied and understood. This is partly because gravity
is the first established universal interaction, but still without a clear correspondence to
other fundamental interactions. 

In this paper we intend to clarify the issue of gravity waves. There is a common belief
that gravity waves are inevitable consequence of Einstein's General Relativity (GR).
The usual argument for their existence relies on the energy loss observed in the binary systems.
In the next three sections we shall prove the opposite, i.e. the nonexistence of gravity
waves, using the local conservation equations.

Further sections will discuss wave equations, similarities and distinctions between fundamental
interactions, and their phenomenology. Our reasoning and results ars summarized in the last
section.

\section{Local conservation equation of Landau and Lifshitz in GR}

There are a few ways to derive the field equations of the GR found in the literature
\cite{Landau,Weinberg}
(with our convention \cite{Landau}:$\eta_{ij}=diag(+1,-1,-1,-1)$):

\begin{eqnarray}
G_{ij} &=& \kappa T_{ij}, \hspace{60 mm} \\
G_{ij} &\equiv& R_{ij} - \frac{1}{2} R g_{ij},\ \kappa\equiv 8 \pi G_{N} c^{-4}.
\nonumber
\end{eqnarray}

However, it is important to stress the unavoidable presence of the integrability conditions
(Bianchi identities) valid for the tensors of the field equations:

\begin{eqnarray}
\nabla _{k}G^{\ k}_{i . }&=&\nabla _{k}T^{\ k}_{i . }=0, \hspace{50 mm}\\
\nabla _{i}G^{\ k}_{j .}&\equiv &\partial_{i}G^{\ k}_{j .}+\Gamma^{k}_{im}G^{\ m}_{j .}
-\Gamma^{m}_{ij}G^{\ k}_{m .},  \nonumber \\
\Gamma^{k}_{im}&\equiv&\left\{ \begin{array}{c} k \\ i\ m \end{array} \right\}=Christoffel\ symbol. \nonumber
\end{eqnarray}

Since the Bianchi identites represent conservation equations in the curved
spacetime, it could be physically advantageous to search for the local conservation
equations with ordinary partial derivatives. Landau and  Lifshitz \cite{Landau}
accomplished this task using both GR field equations and Bianchi identities:

\begin{eqnarray}
&&\frac{\partial}{\partial x^{k}}[(-g)(T^{ik} + t^{ik})] = 0, \hspace{60 mm}\\
 g &=& det(g_{ij}), \nonumber \\
t^{ik}&=& \frac{1}{2\kappa}\{(2\Gamma^{n}_{lm}\Gamma^{p}_{np}-\Gamma^{n}_{lp}\Gamma^{p}_{mn}
-\Gamma^{n}_{ln}\Gamma^{p}_{mp})(g^{il}g^{km}-g^{ik}g^{lm}) \nonumber \\
&& + g^{il}g^{mn}(\Gamma^{k}_{lp}\Gamma^{p}_{mn}+\Gamma^{k}_{mn}\Gamma^{p}_{lp}
-\Gamma^{k}_{np}\Gamma^{p}_{lm}-\Gamma^{k}_{lm}\Gamma^{p}_{np}) \nonumber \\
&& + g^{kl}g^{mn}(\Gamma^{i}_{lp}\Gamma^{p}_{mn}+\Gamma^{i}_{mn}\Gamma^{p}_{lp}
-\Gamma^{i}_{np}\Gamma^{p}_{lm}-\Gamma^{i}_{lm}\Gamma^{p}_{np}) \nonumber \\
&& + g^{lm}g^{np}(\Gamma^{i}_{ln}\Gamma^{k}_{mp}-\Gamma^{i}_{lm}\Gamma^{k}_{np})\}.
\nonumber
\end{eqnarray}

This remarkable exact equation relates ordinary derivatives of the quantites 
$(-g)T_{ij}$ and $(-g)t_{ij}$ that are not tensors with respect 
to the general coordinate transformations (even $g$ is not a scalar, but only a scalar density),
but are tensors with respect to the Lorentz transformations. The immediate consequence of
this equation is the existence of the conserved quantities \cite{Landau}:

\begin{eqnarray}
P^{i}&=&P^{i}(T)+P^{i}(t)=const., \hspace{40 mm} \\
P^{i}(T)&\equiv & c^{-1}\int (-g)T^{i0}dV,\ P^{i}(t)\equiv c^{-1}\int (-g)t^{i0}dV.
\nonumber
\end{eqnarray}

The weak field version of this local conservation equation is derived in Weinberg's textbook 
\cite{Weinberg}.

The local conservation equation is crucial for discussions on relativistic astrophysical
systems.

\section{Local conservation equation in the Einstein-Cartan gravity}

The inclusion of the rotational degrees of freedom into the relativistic theory of gravity 
leads to the Einstein-Cartan (EC) theory of gravity \cite{Sciama,Schouten}. We wish to explore whether
the EC gravity implies a similar local conservation equation as GR. Below we present
field and conservation equations of the EC gravity \cite{Hehl}:

\begin{eqnarray}
G^{ij} &=& \kappa (\sigma^{ij}+\stackrel{*}{\nabla_{k}}(\tau^{ijk}-\tau^{jki}+\tau^{kij})), \\
T^{ijk} &=& \kappa \tau^{ijk}, \\
\stackrel{+}{\nabla_{j}}G_{i .}^{\ j} &=& T_{jk .}^{\ \ l} R_{il . .}^{\ \ jk}, \\
\Gamma^{k}_{ij}&=&\left\{ \begin{array}{c} k \\ i\ j \end{array} \right\}+S_{ij . }^{\ \ k}
-S_{j . i}^{\ k\ }+S_{. ij}^{k}, \nonumber \\
T_{ij .}^{\ \ k} &=& S_{ij .}^{\ \ k} + \delta_{i}^{k}S_{jl .}^{\ \ l}-\delta_{j}^{k}S_{il . }^{\ \ l},
\nonumber \\
\stackrel{*}{\nabla_{k}}\tau^{ijk}&\equiv& \nabla_{k}\tau^{ijk}+2 S_{kl .}^{\ \ l}\tau^{ijk}, \nonumber \\
\stackrel{+}{\nabla_{i}}\psi_{j}&\equiv&\nabla_{i}\psi_{j} + 4 S_{i(j .}^{\ \ k}\psi_{k)},\ (ij)\equiv 
\frac{1}{2}(ij+ji), \nonumber \\
\sigma^{ij}&=& 2 [det(-g_{ij})]^{-1/2}\frac{\delta {\cal L}}{\delta g_{ij}},\ 
S_{ij . }^{\ \ k}=torsion\ tensor, \nonumber \\
\tau^{ijk}&=&spin-angular\ momentum\ tensor. \nonumber
\end{eqnarray}

Owing to a solely algebraic relation between spin-angular momentum of matter and torsion of
spacetime, it is possible to write the field equation with the effective energy-momentum
tensor that contains relevant terms with the spin-angular momentum tensor:

\begin{eqnarray}
G^{ij}(\Gamma=\{\ \}) &=& \kappa T_{EC}^{ij}, \hspace{60 mm} \\
T_{EC}^{ij} &=& \sigma^{ij}(\tau) - \kappa F^{ij} (\tau), \nonumber \\
F^{ij}(\tau) &=& 4\tau^{ik}_{..[l}\tau^{jl}_{..k]}+2\tau^{ikl}\tau^{j}_{.kl}
  -\tau^{kli}\tau^{\ \ j}_{kl.} \nonumber \\
& & -\frac{1}{2}g^{ij}(4\tau^{\ k\ }_{m.[l}\tau^{ml}_{..k]}+\tau^{mkl}
  \tau_{mkl}), \nonumber \\
\sigma^{ij}&=&\sigma^{ji},\ F^{ij}=F^{ji},\ [ij]\equiv \frac{1}{2}(ij-ji).
\nonumber
\end{eqnarray}

Using the Bianchi identity for the Einstein's tensor in Riemannian spacetime, the local conservation
equation in the EC gravity follows immediately:

\begin{equation}
\frac{\partial}{\partial x^{k}}[(-g)(T^{ik}_{EC} + t^{ik})] = 0.
\end{equation}

Thus, if the EC gravity appears to be the proven theory of gravity, one can apply the
local conservation equation whenever it is necessary.

\section{Binary systems}

The discovery of the binary pulsar B1913+16 by Hulse and Taylor
\cite{Hulse} represents a milestone in astrophysics because
relativistic binary pulsar systems are perfect laboratories
to study general relativity.
In the past decades, very detailed calculations were performed
in the post-Newtonian approximation, with all possible relativity
corrections to measurables (see, for example, \cite{Blandford},
 \cite{Damour}, \cite{Taylor} and references therein).

However, it seems that one important part of the calculations is
not included into the analysis of a relativistic binary 
system.
Namely, the backreaction of the spacetime curvature on the 
observables of a binary bound system is not elucidated 
properly.

The metric can be defined in the following form \cite{Landau}:

\begin{eqnarray*}
g_{ij} = \eta_{ij} + h_{ij} .
\end{eqnarray*}

It is suitable for the treatment of an isolated bound system if
we assume that $h_{ij}$ vanishes at infinity.

The calculation of the energy flow in the direction of the $x$ axis
shows the following [note that: $h_{ij}(t-x/c)$],\cite{Landau}:

\begin{eqnarray*}
t^{01}=\frac{1}{2\kappa} [(\frac{\partial h_{22}}{\partial t})^{2}
+(\frac{\partial h_{23}}{\partial t})^{2}],\ 
T^{01}=-t^{01} + \frac{1}{\kappa} \frac{\partial}{\partial x}
[h_{22} \frac{\partial h_{22}}{\partial x}
+h_{23} \frac{\partial h_{23}}{\partial x}].
\end{eqnarray*}

The term with the total derivatives vanishes after integration, not affecting
the local conservation equation.
The components $h_{ij}$ are proportional to the second partial derivative in 
the time variable of
the quadrupole moment of the system $h_{ij} \propto \frac{\partial^{2} D_{ij}}
{\partial t^{2}}$ \cite{Landau}.
The averaging procedure decribed in \cite{Landau}
leads to the well known energy loss of the binary systems \cite{Peters}.

Thus, the orbital period change rate due to the "curvature backreaction"
is, after averaging over one period of
the motion 
\cite{Blandford,Peters,Landau}:

\begin{eqnarray*}
\dot{P}_{b} = f(P_{b},e,m_{p},m_{c}),
\end{eqnarray*}
\begin{eqnarray}
f(P_{b},e,m_{p},m_{c})&=&-\frac{192\pi G_{N}^{5/3}}{5}
(\frac{P_{b}}{2\pi})^{-5/3}(1-e^{2})^{-7/2} \nonumber \\
& &\times(1+\frac{73}{24}e^{2}+\frac{37}{96}e^{4})
m_{p}m_{c}(m_{p}+m_{c})^{-1/3}, \nonumber
\end{eqnarray}
\begin{eqnarray*}
e=eccentricity,\ m_{p;c}=mass\ of\ the\ pulsar;companion.
\end{eqnarray*}

However, we now know from the local Landau-Lifshitz conservation equation that the kinetic energy loss 
is compensated by the potential energy gain (Eq.(4)):

\begin{eqnarray*}
\delta E(T^{ik};kinetic\ energy) + \delta E(t^{ik};potential\ energy) = 0.
\end{eqnarray*}

Small changes of the quadrupole gravity potentials cannot substantially
influence the calculation of the quadrupole moments of the system.

Evidently, there is no energy for gravitational waves. The energy loss is evaluated in the center of mass
Lorentz system. We can evaluate both the kinetic energy loss and
the potential energy gain in some other Lorentz system, but according to the local conservation of energy,
their sum always vanishes.

\section{Wave equation}

The waves in theoretical physics represent the solutions of the wave equations, i.e. the equations
that describe the time evolution of waves in the Minkowski spacetime. For example, the field equations
in GR (or EC gravity) could be represented in a form similar to the wave equations; see 
 ref. \cite{Thorne}, where in his notation Eq.(5.2b) has the following form:

\begin{eqnarray}
(-\partial_{t}^{2}+\nabla^{2})\bar{h}^{\alpha\beta}&=&-16\pi \tau^{\alpha\beta}, 
\hspace{50 mm} \\
\tau^{\alpha\beta}&=&(-g)(T^{\alpha\beta}+t^{\alpha\beta}_{LL})
+(16\pi)^{-1}[\bar{h}^{\alpha\mu}_{\ \ ,\nu}\bar{h}^{\beta\nu}_{\ \ ,\mu}
-\bar{h}^{\alpha\beta}_{\ \ ,\mu\nu}\bar{h}^{\mu\nu}],  \nonumber \\
\bar{h}_{jk}(x)&=&16 \pi \int G^{+}_{jk\cdot pq}(x,x') \tau_{pq}(x')d^{4}x'. \nonumber
\end{eqnarray}

Obviously, this is an integral equation for $\bar{h}_{jk}$, not the wave equation (see also \cite{Sciama1}).
One can perform the weak field approximation ending in the wave equation but the resulting
equation does not belong to the theory invariant under the general coordinate transformations \cite{Weinberg};
Eq.(10.1.10) in Weinberg's notation:

\begin{eqnarray*}
(\nabla^{2}-\frac{\partial^{2}}{\partial t^{2}})h_{\mu\nu}=-16 \pi G S_{\mu\nu}.
\end{eqnarray*}

However, we have to affiliate to the approximate field equation the corresponding approximate Bianchi
identity. We know from the preceding sections that the curvature feedback is always 
present, owing to the Landau-Lifshitz conservation equation, preventing the propagation of the energy 
in the form of the tensor wave.

\section{GR and EC gravity vs. intrinsic local gauge interactions}

The electromagnetic, weak and strong interactions differ substantially from the GR because they
are confirmed to be the intrinsic unitary local gauge symmetries. On the other hand, the GR (or the EC gravity)
are based on spacetime symmetries that cannot propagate in the form of local tensor waves,
because of the integrability conditions (Bianchi identities).

Conformal and discrete symmetries play somewhat different roles in the nonsingular and causal
$SU(3)\times SU(2)\times U(1)$ BY theory of ref. \cite{Palle1,New1,New2} than in the nonsingular EC gravity.

The six dimensional conformal space of the conformal symmetry is perfectly suitable for the 
conformal $SU(3)$ unification of all the observed $SU(3)\times SU(2)\times U(1)$ local gauge
forces \cite{Palle1}: $8\times 6 = 8\times 4 + 3\times 4 + 4=gauge\ degrees\ of\ freedom$.
Trace anomaly (or the divergence of the dilatational current) and Wick's theorem help us 
in generating the elementary particle masses \cite{Palle1}.

The violation of the discrete symmetries by weak interactions is strongly connected 
with the $SU(2)$ global anomaly and
the topology of the unitary symmetries in the conformal space \cite{Palle1}.

On the other hand, the four dimensional spacetime is the lowest dimensional spacetime for 
which the conformal (Weyl) tensor does not vanish \cite{Weinberg}.
The conformal tensor plays a crucial role when fixing the cosmic mass density within the
EC gravity \cite{Palle2,Davor}. The chirality of the vorticity of the Universe appears
to be closely related to the chirality of the weak interactions \cite{Palle3,New3,New4}.
The absence of singularity in the EC gravity, because of the presence of 
spin densities \cite{Trautman,Palle2}, matches the scale of the weak interactions \cite{Palle1}.

Evidently, the roles of conformal and discrete symmetries in the relativistic theories of gravity and 
intrinsic local gauge theories are important, but their manifestations are different according to the
substantially diverse nature of the force of gravity compared to the electroweak and strong forces.

\section{Indirect evidence of gravity waves or just local conservation of energy-momentum}

We know that in many binary systems
the kinetic energy loss is calculated and observed to very high precision.  

For example, very precise measurements of the binary pulsar B1913+16 with
an overdetermined set of measurables give \cite{Weisberg}

\begin{eqnarray*}
\dot{P}_{b}(gen.\ rel.)=f(P_{b},e,m_{p},m_{c})=
(-2.40247\pm 0.00002)\times 10^{-12},
\end{eqnarray*} 
\begin{eqnarray*}
\dot{P}_{b}(measured)=(-2.4086\pm 0.0052)\times 10^{-12}.
\end{eqnarray*}

The kinetic energy loss,
which is due to the nonvanishing orbital period change rate, is 
compensated by the potential energy gains hidden in $h_{ij}$
according to the Landau-Lifshitz local conservation equation:

\begin{eqnarray*}
\Delta E (kin.\ en.,\dot{P}_{b}\neq 0) +
\Delta E (pot.\ en.\ change\ in\ h_{ij}) = 0 .
\end{eqnarray*}

If one neglects the inevitable change of the potential energy,
the kinetic energy loss in $\dot{P}_{b}$ can be compensated
by the introduction of the energy of the hypothetical emission of the
tensor field (gravity waves):

\begin{eqnarray*}
\Delta E (kin.\ en.,\dot{P}_{b}\neq 0) +
\Delta E (en.\ deposited\ in\ grav.\ waves) = 0.
\end{eqnarray*}

However, it is possible to search for the change rate of the quadrupole potentials
of binary systems by gravitational lensing, by measuring the change of the polarization 
of photons passing nearby \cite{Palle4}. 

We showed recently that the pulsar timing array data can be interpreted 
as a vortical motions of pulsar's photons within rotating and expanding
Universe without any reference to the stochastic gravitational wave
background \cite{Palle5}.

LISA observatory could be the best oportunity to measure the anomalous acceleration
in the solar system which is a inevitable consequence of the cosmic acceleration \cite{New5}.

\section{LIGO detections: gravity waves or geophysical phenomenon}

Starting from 2016 the LIGO collaboration reported a great number of events 
accepted and described as gravity waves from the binary systems with 
merging black holes, 
neutron stars, etc. \cite{LIGO}. However, to our knowledge, only one LIGO event 
GW 170817 is correlated to some GRB event, namely, GRB 170817A. It is not clear
which event is reported first, GRB of the Fermi satellite or GW of the LIGO.
After establishing the optical, radio, infrared, X-ray, gamma ray and neutrino
follow up network, no corrrelations of LIGO GW events are found with any kind of
astrophysical sources. Some LIGO GW events fitted with black hole binary system
collapse demand the theoretically forbidden mass of the black hole. 

One month after the announcement of the first GW event, the alternative
interpretation of the LIGO events was formulated \cite{Palle6}, communicated
to the LIGO scientists and later cited in the book of H. Collins \cite{Collins}.

We repeat the key argument of ref. \cite{Palle6} that the ocean tidal bulges
are responsible for the GW events at LIGO.
The order of the magnitude estimate of the ocean bulge height can be done assuming the work per unit volume
of the static fluid before and after the action of the conservative force of gravity:  

\begin{eqnarray}
\rho (g_{Earth} - g_{Moon})(h + \Delta h) = \rho g_{Earth} h, \nonumber
\end{eqnarray}

where $\rho$ is the mass density of the fluid, $g_{Earth}=9.81 ms^{-2}$, $g_{Moon}=G_{N}\frac{M_{Moon}}{d^{2}}$,
where $d$ is the Moon to Earth distance, $h$ is the average depth of the ocean. It follows:

\begin{eqnarray}
\Delta h = h \frac{g_{Moon}}{g_{Earth} - g_{Moon}}. \nonumber
\end{eqnarray}

The average depth of the Indian ocean is $h \simeq 4 km$. However, the Indian ocean trenches have up to $h \simeq 7 km$ depth
($\Delta h = {\cal O}(10 cm)$).
Hence, the moving tidal bulges formed with the maximal Moon's and Sun's gravity
forces when crossing the ocean trenches are producing the maximal force on the LIGO
detectors.

Precisely this effect is observed in the data analysis of few LIGO GW events 
by the astrophysical group at the Niels Bohr Institute in Copenhagen
\cite{Hao}. Namely, they found the correlations between the GW events and
background time lags.

Sapienti sat.

\end{document}